\pdfoutput=1
\documentclass[preprint,showpacs,pra]{revtex4}
\usepackage{amsfonts}
\usepackage{amssymb}
\usepackage{amsmath}
\usepackage{graphicx}
\usepackage{caption}
\usepackage{subcaption}

\begin{document}

\title{The use of coherent states in the excitation and ionization of atoms
by a short symmetric pulse.}
\date{\today }

\begin{abstract}
We propose a new model for describing the excitation of atoms by short
symmetric pulses. The theory is based on the impulse approximation and
incorporates a confining potential approximated by a harmonic oscillator.
Results are presented for a wide variety of parameters, including field
intensity and laser frequency. Permitted and forbidden transitions are
analyzed in detail. The photon energy spectrum presents peaks attributed to
multiphoton absorption. Our model demonstrates robustness and reliability,
paving the way for addressing more complex scenarios.
\end{abstract}

\author{J. E. Miraglia}
\maketitle
\affiliation{Instituto Astronomía y Física del Espacio.
Consejo Nacional de Investigaciones Científicas y Técnicas. 
Universidad de Buenos Aires. Argentina.}

\section{INTRODUCTION}

The interaction of coherent electromagnetic radiation with atoms has become
an extraordinarily active area of research, both experimentally and
theoretically \cite{Vrakking2024}. On the theoretical front, most efforts
have focused on the full numerical solution of the time-dependent Schr\"{o}%
dinger equation\ (TDSE) on a grid. This approach is highly successful for
single-electron systems. However, more complex targets demand a deeper
understanding of the physical processes involved, and distorted-wave methods
serve as valuable tools for exploration.

In this study, we develop a new theoretical model to describe inelastic
transitions induced by a short symmetric pulse. It is based on the
double-distorted rigid traveling orbital (RTO) approximation, related to the
so-called impulse approximation\ \cite{Macri03, Gravielle2012}. This
approximation incorporates the effects of the laser field on both the
initial and final states on equal footing. The impulse approximation is a
well-established theory applicable to energetic pulses, such as those in the
UV region. However, for low-energy lasers and/or weak electric field
intensities, the assumption that the electric field overwhelmingly dominates
the interaction with the nucleus --allowing the electron to move exclusively
under its influence-- does not hold. To address this limitation, we
introduce a confining potential. This potential ensures dominance near the
nucleus, allowing the electric field to take control at greater distances.
The confining potential in the initial channel is approximated by a harmonic
oscillator, enabling us to construct a double-distorted wave referred to as
the Harmonic Oscillator (HO) model. This approach draws inspiration from the
seminal work of Kerner, published in 1958 \cite{Kerner58}. The inclusion of
a confining potential causes the electron to end up in a coherent state, a
subject that has recently received attention \cite{Jia24}. We analyze the
energy of these states, which appears to be an attractive and largely
unexplored topic. We compare the RTO\ and HO models under various
conditions, including changes in laser frequency (ranging from infrared to
UV) and field strength. Permitted transitions in hydrogen (1s $\rightarrow $
np), as well as forbidden transitions (1s $\rightarrow $ ns, 1s $\rightarrow 
$ 3d, and 1s $\rightarrow $ 4f), are examined and discussed. For ionization,
comparison with the TDSE is made. This paper is organized as follows:
Section 2 develops the two theories, RTO and HO; Section 3 presents the
results; and Section 4 outlines the conclusions and future directions.
Atomic units are used throughout.

\section{THEORY}

\subsection{The electric field}

Throughout this article, we use the usual expression for the electric field%
\begin{equation}
\overrightarrow{E}(t)=\widehat{z}E_{0}\sin (\omega t+\phi )\sin ^{2}\left(
\pi t/\tau \right) ,  \label{1}
\end{equation}%
where $\omega $ is the photon energy, ranging from 0.05 to 2. This range
covers wavelengths from infrared lasers ($\omega _{IR}=0.057$, corresponding
to 800 nanometers) to the extreme ultraviolet region ($\omega =55$ electron
volts), which is far beyond the ionization energy. The pulse duration, $\tau 
$, is defined through the number of cycles, $N_{cy}$, such that $\tau =2\pi
N_{cy}/\omega $. In our case, $N_{cy}=4$, and so $\tau $ ranges from 12.5
a.u. (303 attoseconds) to 502 a.u. (12 femtoseconds). The electric field
strength $E_{0}$ is varied from $0.01\ (10^{12}\,W/cm^{2})$ to $1.0\
(10^{16}\,W/cm^{2})$, spanning quiver amplitudes, $\alpha _{\max
}=E_{0}/\omega ^{2}$, from 0.0025 to 400.

Using this definition of $\overrightarrow{E}(t)$, the vector potential $%
\overrightarrow{A}(t)=\int_{0}^{t}dt^{\prime }\overrightarrow{E}(t^{\prime
}),$ satisfies $\overrightarrow{A}(\tau )=0$, regardless of the phase shift $%
\phi $. The corresponding quiver amplitude, $\alpha
(t)=\int_{0}^{t}dt^{\prime }\overrightarrow{A}(t^{\prime })$, results in $%
\alpha (t=\tau )=0$ only when $\phi =\pi /2$, which corresponds to a purely 
\textit{symmetric} pulse. For any other phase shift, i.e., $\phi \neq \pi /2$%
, we find that $\alpha (\tau )\neq 0$. Specifically, for $\phi =0$ or $\pi $%
, the pulse is \textit{asymmetric}. This work deals with \textit{symmetric}
pulses.

\subsection{The rigid traveling orbital}

Consider the following rigid traveling orbital (RTO) initial state%
\begin{equation}
\chi _{i}^{RTO+}(\overrightarrow{q}^{+},\overrightarrow{p}^{+}|%
\overrightarrow{r},t)=\psi _{i}(\overrightarrow{r}-\overrightarrow{q}%
^{+})\exp [i\overrightarrow{p}^{+}\cdot \overrightarrow{r}\
-iL^{+}-iE_{i}t]\ ,  \label{10}
\end{equation}%
where $\psi _{i}$ is the initial electronic state, satisfying $H_{0}\psi
_{i}=E_{i}\psi _{i}$, with $H_{0}=-%
{\frac12}%
\nabla _{r}^{2}+V_{C}(r)$ the undistorted Hamiltonian, and $V_{C}(r)$ the
central potential ($V_{C}(r)=-1/r$ in our case). The real magnitude $%
L^{+}=L^{+}(t)$ will be identified below as the ponderomotive energy. The
so-defined traveling orbital can also be called a \textit{coherent state};
in this case, the real quantities $\overrightarrow{q}^{+}(t)$ and $%
\overrightarrow{p}^{+}(t)$ are called \textit{labels} characterizing the
state. For brevity, we will most often denote them simply as $L^{+},\ 
\overrightarrow{q}^{+}$, and $\overrightarrow{p}^{+}$ when possible, though
one should always keep in mind that these are time-dependent quantities. The
rigid traveling state $\chi _{i}^{\text{RTO}+}$ is \textit{normalized} at
all times, i.e.\ $\left\langle \chi _{i}^{\text{RTO}+}|\chi _{i}^{\text{RTO}%
+}\right\rangle =1.$ Regardless of the values of $\overrightarrow{q}^{+}$
and $\overrightarrow{p}^{+}$, the \textit{continuity equation} imposes%
\begin{equation}
\overrightarrow{p}^{+}(t)=\frac{d}{dt}\overrightarrow{q}^{+}(t).  \label{30}
\end{equation}%
The labels $\overrightarrow{q}^{+}$ and $\overrightarrow{p}^{+}$ are related
to the mean values $\overrightarrow{q}^{+}=\left\langle \chi _{i}^{\text{RTO}%
+}\left\vert \overrightarrow{r}\right\vert \chi _{i}^{\text{RTO}%
+}\right\rangle $ and $\overrightarrow{p}^{+}=\left\langle \chi _{i}^{\text{%
RTO}+}\left\vert \frac{1}{i}\overrightarrow{\nabla }_{\overrightarrow{r}%
}\right\vert \chi _{i}^{\text{RTO}+}\right\rangle ,$ satisfying the
relations of coherent states as introduced by Klauder \cite{Klauder63}.

In the presence of the electric pulse, the total Hamiltonian $H$ reads $\
H=H_{0}+\overrightarrow{E}(t)\cdot \overrightarrow{r},$ where the electric
pulse $\overrightarrow{E}(t)$ starts at $t=0$ and ends at $t=\tau $ as in
Eq.(\ref{1}). The initial condition imposes $\chi _{i}^{RTO+}(t=0)=\psi _{i}(%
\overrightarrow{r})\exp [-iE_{i}t]\ ,$ or equivalently%
\begin{equation}
\overrightarrow{q}^{+}(0)=\overrightarrow{p}^{+}(0)=0.  \label{55}
\end{equation}%
Applying $H-i\frac{\partial }{\partial t}$ to the state $\chi _{i}^{\text{RTO%
}+}$ produces the remaining initial potential $W_{i}^{\text{RTO}+}$, i.e.%
\begin{equation}
\left( H-i\frac{\partial }{\partial t}\right) \chi _{i}^{\text{RTO}%
+}(r,t)=W_{i}^{\text{RTO}+}\chi _{i}^{\text{RTO}+}(r,t),  \label{60}
\end{equation}%
\begin{equation}
W_{i}^{\text{RTO}+}=V_{C}(\overrightarrow{r})-V_{C}(\overrightarrow{r}-%
\overrightarrow{q}^{+})+\overrightarrow{r}\cdot \left[ \frac{d^{2}}{dt^{2}}%
\overrightarrow{q}^{+}-e\ \overrightarrow{E}(t)\right] -\left[ \frac{d}{dt}%
L^{+}(t)]-\frac{1}{2}\left\vert \overrightarrow{p}^{+}\right\vert ^{2}\right]
.\   \label{69}
\end{equation}%
Neglecting the first squared bracket of Eq.(\ref{69}), we arrive at the
Newton equation%
\begin{equation}
\frac{d}{dt}\overrightarrow{q}^{+}(t)=\overrightarrow{p}^{+}(t)=%
\int_{0}^{t}dt^{\prime }\ e\overrightarrow{E}(t^{\prime })  \label{90}
\end{equation}%
which describes the motion of a \textit{central} \textit{classical} particle
of charge $e=-1$ in the electric field $\overrightarrow{E}(t).\ $We call it 
\textit{central} particle because it describes the motion of the center of
the electron density, which, in this model, maintains its \textit{rigid}
form. We can now clearly identify that $\overrightarrow{p}^{+}(t)=-%
\overrightarrow{A}(t).\ $\ Neglecting the second squared bracket of Eq.(\ref%
{69}), we find the expression of the ponderomotive term 
\begin{equation}
L^{+}(t)=\int_{0}^{t}dt^{\prime }\frac{1}{2}\left\vert \overrightarrow{p}%
^{+}(t^{\prime })\right\vert ^{2}.  \label{95}
\end{equation}%
The remaining initial potential $W_{i}^{\text{RTO}+}$,%
\begin{equation}
W_{i}^{\text{RTO}+}=V_{C}(\overrightarrow{r})-V_{C}(\overrightarrow{r}-%
\overrightarrow{q}^{+}),  \label{70}
\end{equation}%
can be identified as two Coulomb charges: one positive, static, at the
center ($\overrightarrow{r}=0$), and another, negative, moving and centered
at $\overrightarrow{r}=\overrightarrow{q}^{+}(t),$ portraying a dynamic
dipole.

Regardless of the value of $\phi $, the final momentum $\overrightarrow{p}%
(\tau )=0$, but the final position $\overrightarrow{q}^{+}(t=\tau )$ \textit{%
does} depend on $\phi $, and its value is%
\begin{equation}
\overrightarrow{q}^{+}(\tau )=\frac{E_{0}}{\omega ^{2}}\frac{\pi N_{cy}}{%
(N_{cy}^{2}-1)}\cos (\phi ).  \label{110}
\end{equation}%
In other words, in the phase space \{$\overrightarrow{q}^{+},\overrightarrow{%
p}^{+}$\}, the central classical particle starts at $t=0$ at the position $%
\{0,0\}$ and ends at $t=\tau $ at $\{\overrightarrow{q}^{+}(\tau ),0\}$. For 
$\phi =\pi /2$ (symmetric pulse), $\overrightarrow{q}^{+}(\tau )=0$ and so%
\begin{equation}
\chi _{i}^{\text{RTO}+}(t=\tau )=\psi _{i}(\overrightarrow{r})\exp
[-iL^{+}(\tau )-iE_{i}\tau ]\ ,  \label{121}
\end{equation}%
which is essentially the initial state times a constant phase factor $%
L^{+}(\tau )$. In other words, the traveling orbital starts moving at $t=0$
and returns to the origin at $t=\tau .$ Alternatively, $\overrightarrow{q}%
^{+}(\tau )$ $\rightarrow 0,$ as $N_{cy}\rightarrow \infty $ and/or $\omega
\rightarrow \infty .$

By setting $\overrightarrow{q}^{+}=\overrightarrow{p}^{+}=0,$ we arrive at
the simple Born approximation,%
\begin{equation}
\chi _{i}^{\text{RTO}+}(0,0|\overrightarrow{r},t)=\chi _{i}^{\text{B}}(%
\overrightarrow{r},t)=\psi _{i}(\overrightarrow{r})\exp [-iE_{i}t],
\label{130}
\end{equation}%
whose residual potential is simply the electric perturbation $W_{i}^{B}=%
\overrightarrow{E}(t)\cdot \overrightarrow{r}.$

Similar expressions can be obtained for the final-channel traveling orbital
by defining%
\begin{equation}
\chi _{i}^{\text{RTO}-}(\overrightarrow{q}^{-},\overrightarrow{p}^{-}|%
\overrightarrow{r},t)=\psi _{f}(\overrightarrow{r}-\overrightarrow{q}^{-})\
\exp [i\overrightarrow{p}^{-}\cdot \overrightarrow{r}\ -iL^{-}-iE_{f}t],
\label{135}
\end{equation}%
satisfying the outgoing conditions $\chi _{f}^{\text{RTO}-}(\overrightarrow{q%
}^{-},\overrightarrow{p}^{-}|\overrightarrow{r},\tau )=\psi _{f}(%
\overrightarrow{r})\exp [-iE_{f}\tau ]$ \ with a remaining potential%
\begin{equation}
W_{f}^{\text{RTO}-}=V_{C}(\overrightarrow{r})-V_{C}(\overrightarrow{r}-%
\overrightarrow{q}^{-}).  \label{138}
\end{equation}%
At any time, the Newton equation reads%
\begin{equation}
\frac{d}{dt}\overrightarrow{q}^{-}(t)=\overrightarrow{p}^{-}(t)=\int_{\tau
}^{t}dt^{\prime }\ e\overrightarrow{E}(t^{\prime }).  \label{142}
\end{equation}%
with%
\begin{equation}
\overrightarrow{q}^{-}(\tau )=\overrightarrow{p}^{-}(\tau )=0.  \label{140}
\end{equation}

\subsection{The strong electric field approximation}

One can easily write the amplitude of transition using both rigid traveling
orbitals, $\chi _{i}^{\text{RTO}+}$ and $\chi _{f}^{\text{RTO}-},$ in the
initial and final channels, respectively. In atomic collision theory, the
matrix element so developed is sometimes named the impulse approximation \ 
\cite{Macri03}. This model considers the electron to be weakly bound, so it
can be regarded as free, which is equivalent to assuming that the electric
field largely dominates over the central Coulomb potential. In this case, we
define the RTO amplitude $A_{f}^{\text{RTO}}(t)$ as%
\begin{equation}
A_{f}^{\text{RTO}}=\left\langle \chi _{f}^{\text{RTO}-}\left\vert W_{f}^{%
\text{RTO}-}\right\vert \chi _{i}^{\text{RTO}+}\right\rangle .  \label{145}
\end{equation}%
Using the variable $\overrightarrow{R}$=$\overrightarrow{r}-\overrightarrow{q%
}^{-},$ $\overrightarrow{q}=\overrightarrow{q}^{+}-\overrightarrow{q}^{-},$
and $\overrightarrow{p}=\overrightarrow{p}^{+}-\overrightarrow{p}^{-},$ $%
A_{f}^{\text{RTO}}(t)$ reads%
\begin{eqnarray}
A_{f}^{\text{RTO}}(t) &=&\exp [i\Phi _{1}(t)]\int d\overrightarrow{R}\ \psi
_{f}^{\ast }(\overrightarrow{R})\left[ W_{f}^{\text{RTO}-}\right] \psi
_{i}\left( \overrightarrow{R}-\overrightarrow{q}\right) ,\ \   \label{210} \\
\Phi _{1}(t) &=&\left( E_{f}-E_{i}\right) t-\ \left[ L^{+}(t)-L^{-}(t)\right]
,  \label{230}
\end{eqnarray}%
where we have used the property $\overrightarrow{p}=\overrightarrow{p}%
^{+}(t)-\overrightarrow{p}^{-}(t)=\overrightarrow{p}(\tau )=0.$ The
interesting point is that the value of $\overrightarrow{q}(t)$ along the
electric field ($\widehat{z}$) is constant at all times, because%
\begin{equation}
\overrightarrow{q}(t)=\overrightarrow{q}^{+}(t)-\overrightarrow{q}^{-}(t)=\
\int_{0}^{\tau }dt^{\prime }\overrightarrow{A}(t^{\prime })=q^{+}(\tau ),
\label{232}
\end{equation}%
and $\overrightarrow{q}^{+}(\tau )$ is given by Eq.(\ref{110}), which
depends on $\phi .$ In similar fashion, the difference of ponderomotive
energies, $L(t)=L^{+}(t)-L^{-}(t),$ gives rise to a definite integral, also
independent of time, i.e.%
\begin{equation}
L^{+}(t)-L^{-}(t)=\ \int_{0}^{\tau }dt^{\prime }\frac{1}{2}\left\vert 
\overrightarrow{p}^{+}(t^{\prime })\right\vert ^{2}=L^{+}(\tau ),
\label{235}
\end{equation}%
so the ponderomotive energy makes no contribution to the probability.
Therefore, for symmetric pulses, the states are orthonormal, $i.e.\ $ $%
\left\langle \chi _{f}^{\text{RTO}-}|\chi _{i}^{\text{RTO}+}\right\rangle
=\delta _{fi}.$

The probability of transition is defined as $P_{f}^{\text{RTO}}=P_{f}^{\text{%
RTO}}(t=\tau ),$ and it reads%
\begin{equation}
P_{f}^{\text{RTO}}(\tau )=\left\vert \ \ \int_{0}^{\tau }\ dt^{\prime
}A_{f}^{\text{RTO}}(t^{\prime })\right\vert ^{2}.  \label{240}
\end{equation}%
It is very important to note here that since $\overrightarrow{q}^{-}(t>\tau
)=0,$ the perturbation $W_{f}^{\text{RTO}-}=V_{C}(\overrightarrow{R}+%
\overrightarrow{q}^{-})-V_{C}(\overrightarrow{R})=0,$ and consequently $%
A_{f}^{\text{RTO}}(t^{\prime }>\tau )=0$ and $P_{f}^{\text{RTO}}(t>\tau
)=P_{f}^{\text{RTO}}(\tau ).$

For a small perturbation $E(r,t)$ (and consequently a small $\overrightarrow{%
q}^{-}(t)$), one can expand%
\begin{equation}
V_{C}(\overrightarrow{R}+\overrightarrow{q}^{-})-V_{C}(\overrightarrow{R}%
)\approx +\overrightarrow{\nabla }_{\overrightarrow{R}}V_{C}(\overrightarrow{%
R})\cdot \overrightarrow{q}^{-}(t)\ +\ \mathcal{O}(q^{-\ 2}),  \label{250}
\end{equation}%
and using the basic algebra of commutators, such as that used in the
photoelectric effect, we arrive at the Born approximation%
\begin{equation}
A_{f}^{\text{RTO}}(\tau )\underset{E_{0}\rightarrow 0}{\longrightarrow }%
A_{f}^{\text{B}}(\tau )=-\sqrt{2\pi }\ \left\langle \psi _{f}\left\vert
z\right\vert \psi _{i}\right\rangle \ \widetilde{E}(0,\tau |E_{i}-E_{f}),
\label{260}
\end{equation}%
where $\widetilde{E}$ is the incomplete Fourier transform of the electric
field, defined in general as%
\begin{equation}
\widetilde{E}(t_{1},t_{2}|\omega _{0})=\frac{1}{\sqrt{2\pi }}%
\int_{t_{1}}^{t_{2}}dt^{\prime }\ E(t^{\prime })\ \exp [-i\omega
_{0}t^{\prime }],  \label{262}
\end{equation}%
which in our case has a very simple closed form.

The Coulomb-Volkov (CV) approximation \cite{Macri03}\ consists of neglecting 
$\overrightarrow{q}^{-}$ in\ $\chi _{f}^{\text{RTO}-}$, $i.e.$ $\chi _{f}^{%
\text{RTO}-}(0,\overrightarrow{p}^{-}|\overrightarrow{r},\tau )=\chi
_{f}^{CV-}(\overrightarrow{p}^{-}|\overrightarrow{r},\tau ),$ and replacing $%
\chi _{i}^{\text{RTO}+}$ by the first Born wave function $\chi _{i}^{\text{B}%
}$, to give%
\begin{equation}
A_{f}^{\text{CV}}=\left\langle \chi _{f}^{\text{CV}-}\left\vert 
\overrightarrow{E}(t)\cdot \overrightarrow{r}\right\vert \chi _{i}^{\text{B}%
}(\overrightarrow{r},t)\right\rangle .  \label{263}
\end{equation}%
The wave function $\chi _{\overrightarrow{k}}^{CV-}(\overrightarrow{p}^{-}|%
\overrightarrow{r},\tau )$ satisfies neither the continuity equation nor the
Newton equation.

\subsection{The confining potential}

While the RTO has an appealing form, for realistic values of the electric
field, it is not sound for describing the initial bound state, since the
electron is treated as unconstrained by the nucleus. One would expect the
RTO model to hold at very large electric field, $i.e.\ E_{0}>>1.$ But for \ $%
E_{0}\lesssim 1,$ the notion that the electric field can arbitrarily govern
the electron is incorrect, since the electric field competes with the
Coulomb attraction of the nucleus. Thus, the central Coulomb potential plays
a pivotal role in this context. To address this issue, we introduce a
classical confining potential (CP), $V_{\text{CP}}(\overrightarrow{Q}^{+})$,
and its corresponding force $\overrightarrow{F}_{\text{CP}}(\overrightarrow{Q%
}^{+})=-\overrightarrow{\nabla }_{\overrightarrow{Q}}V_{\text{CP}}(%
\overrightarrow{Q}^{+})$, which restricts the movement of the electron. To
incorporate the effect of $V_{\text{CP}}$ into the Hamiltonian, let us add
and subtract the following quantities on the right-hand side of Eq. (\ref{69}%
)

\begin{equation}
\pm \left( \overrightarrow{r}\cdot \overrightarrow{F}_{\text{CP}}(%
\overrightarrow{Q}^{+})+U_{\text{CP}}(\overrightarrow{Q}^{+})\right) ,
\label{270}
\end{equation}%
with%
\begin{equation}
U_{\text{CP}}(\overrightarrow{Q}^{+})=\overrightarrow{F}_{\text{CP}}(%
\overrightarrow{Q}^{+})\cdot \overrightarrow{Q}^{+}(t)+V_{\text{CP}}(%
\overrightarrow{Q}^{+}),  \label{272}
\end{equation}%
Note that a new central (classical) position variable $\overrightarrow{Q}%
^{+}=\overrightarrow{Q}^{+}(t)$ is used instead of $\overrightarrow{q}^{+}.$
After simple algebra, the classical Newton equation (\ref{90}) now reads%
\begin{equation}
\frac{d^{2}}{dt^{2}}\overrightarrow{Q}^{+}(t)=e\ \overrightarrow{E}(t)+%
\overrightarrow{F}_{\text{CP}}\left( \overrightarrow{Q}^{+}(t)\right) =\frac{%
d}{dt}\overrightarrow{P}^{+}(t),  \label{280}
\end{equation}%
which describes a classical particle moving under the influence of the
electric field $\overrightarrow{E}(t)$ \textit{and} the confining potential $%
V_{\text{CP}}(\overrightarrow{Q}^{+})$. \ This potential now imposes a new
trajectory for the central particle in the phase space \{$\overrightarrow{Q}%
^{+},\overrightarrow{P}^{+}$\}. The magnitude $L^{+}$ absorbs $U_{\text{CP}}(%
\overrightarrow{Q}^{+})$, thereby transforming $L^{+}$ into $L_{\text{CP}%
}^{+}$, where%
\begin{equation}
\ L_{\text{CP}}^{+}(t)=\int_{0}^{t}dt^{\prime }\left[ \frac{1}{2}\left\vert 
\overrightarrow{P}^{+}(t^{\prime })\right\vert ^{2}+U_{\text{CP}}(%
\overrightarrow{Q}^{+})\right] .  \label{290}
\end{equation}%
Four terms were introduced in (\ref{270}): one was absorbed into the Newton
equation, the other into the term $L_{\text{CP}}^{+},$ and the two remaining
terms stay in the remaining perturbation to produce $W_{i}^{\text{CP}+}$.\
The wave function $\chi _{i}^{\text{CP}+}$ retains the same structure as
before%
\begin{equation}
\chi _{i}^{\text{CP}+}(\overrightarrow{r},t)=\psi _{i}(\overrightarrow{r}-%
\overrightarrow{Q}^{+})\ \exp [i\overrightarrow{P}^{+}\cdot \overrightarrow{r%
}\ -iL_{\text{CP}}^{+}-iE_{i}t].  \label{312}
\end{equation}
Note that $\overrightarrow{q}^{+}$ solves Eq.(\ref{90}) while $%
\overrightarrow{Q}^{+}$ solves Eq.(\ref{280})\ including the confining
potential. The reasoning behind introducing the terms in Eq.(\ref{272})
becomes evident when calculating the mean value%
\begin{eqnarray}
\left\langle H_{i}^{\text{CP+}}\right\rangle &=&\left\langle \psi _{f}^{%
\text{CP+}}\left\vert i\frac{\partial }{\partial t}\right\vert \ \psi _{i}^{%
\text{CP+}}\right\rangle ,  \label{320} \\
&=&E_{i}+\frac{1}{2}\left\vert \overrightarrow{P}^{+}\right\vert ^{^{2}}+\
V_{\text{CP}}(\overrightarrow{Q}^{+})\ +\overrightarrow{E}(t)\cdot 
\overrightarrow{Q}^{+},\ \ \   \label{330}
\end{eqnarray}%
which expresses the correct balance of energy. For $t\leq 0,$ $%
\overrightarrow{P}^{+}=\overrightarrow{Q}^{+}$ $=\overrightarrow{E}=0$, and
the energy is $\left\langle H_{i}^{\text{CP+}}\right\rangle =$ $E_{i},$ the
energy of the initial eigenstate of the unperturbed Hamiltonian. Here it is
important to note that Eq.(\ref{330}) requires $V_{\text{CP}}(%
\overrightarrow{Q}^{+}=0)=0$ in order to reference the energy to that of the
initial state $E_{i}$. For $\tau \geq t\geq 0,$ the electric field transfers
energy (or subtracts it, depending on whether $\overrightarrow{E}$ is
parallel or antiparallel to $\overrightarrow{Q}^{+}$). Finally, for $t\geq
\tau ,\ \overrightarrow{E}=0$, $\overrightarrow{P}^{+}(t)$ and $%
\overrightarrow{Q}^{+}(t)$ oscillate with time such that $\left\langle
H_{i}^{\text{CP+}}\right\rangle $ reduces to the classical Hamiltonian%
\begin{equation}
\left\langle H_{i}^{\text{CP+}}\right\rangle =E_{i}+\frac{1}{2}\left\vert 
\overrightarrow{P}^{+}\right\vert ^{^{2}}+V_{\text{CP}}(\overrightarrow{Q}%
^{+}),\text{ \ \ for }t\geq \tau .  \label{332}
\end{equation}%
adding a constant to the initial energy $E_{i}$. This is the energy of the
coherent state. In principle, this state persists indefinitely, as is
characteristic of a coherent state.

\subsection{The harmonic oscillator}

Given the symmetry of our problem, it is convenient to express the positions
in cylindrical coordinates. Defining the direction of the electric field
along the $\widehat{z}$ axis, $\overrightarrow{E}(t)=E(t)\widehat{z}$, we
can exploit the axial symmetry. In this coordinate system, the position of
the electron is represented as $\overrightarrow{r}=(\rho ,z,\phi )$, while
that of the classical central particle is given by $\overrightarrow{Q}$=($%
Q_{\rho },Q_{z},\phi _{Q})$. The axial symmetry introduces the selection
rule $\Delta m=0$, implying $m_{i}=m_{f}$, though $\Delta l=\pm 1$ is not
enforced. To simplify the problem, the motion of the quasiparticle can be
confined along the field direction $\widehat{z}$, which makes only the $%
Q_{z} $ component relevant.

We have studied various one-dimensional potentials, such as the one proposed
by Clark \cite{Clark92}, which is particularly well suited to our case.
However, solving the Newton equation for this potential requires numerical
methods. In light of this, we resort to the widely applicable
one-dimensional harmonic oscillator (HO):%
\begin{equation}
V_{\text{CP}}(\overrightarrow{Q}^{+})=V_{\text{HO}}(Q_{z}^{+})=\frac{1}{2}%
\omega _{0}^{2}\ Q_{z}^{+\ 2}.  \label{500}
\end{equation}%
satisfying $V_{\text{HO}}(0)=0,$ as required, and the corresponding
confining force is $\overrightarrow{F}_{\text{HO}}(\overrightarrow{Q}%
^{+})=-\omega _{0}^{2}\ Q_{z}^{+}.$ The advantage of the HO is that the
solution of the Newton equations including the electric field has a very
simple expression (see Appendix 1). Note that the magnitude $U_{\text{HO}}$
now reads%
\begin{equation}
U_{\text{HO}}(\overrightarrow{Q}^{+})=\overrightarrow{F}_{\text{HO}}(%
\overrightarrow{Q}^{+})\cdot \overrightarrow{Q}^{+}+V_{\text{HO}}(%
\overrightarrow{Q}^{+}(t)=-V_{\text{HO}}(\overrightarrow{Q}^{+}),
\label{510}
\end{equation}%
and consequently, $L_{\text{HO}}^{+}(t)=\int_{0}^{t}dt^{\prime }\mathcal{L}%
(t^{\prime }),$ where $\mathcal{L}(t^{\prime })$ is the Lagrangian%
\begin{equation}
\mathcal{L}(t^{\prime })=\frac{1}{2}\overrightarrow{P}^{+2}-\frac{1}{2}%
\omega _{0}^{2}\ \overrightarrow{Q}^{+\ 2}.  \label{520}
\end{equation}%
The role of the Lagrangian $\mathcal{L}(t^{\prime })$ in the solution of the
harmonic oscillator in an electric field was already noted by Kerner \cite%
{Kerner58} a long time ago.

Still, the use of $V_{\text{HO}}$ as the confining potential is not yet
physical, because it makes the electron density oscillate rigidly,
independent of the distance to the nucleus. We can remedy this deficiency by
re-defining the Hooke constant to account for the proximity of the nucleus,
setting $\omega _{0}^{2}=\omega _{0}^{2}(\rho ),$ and so we work with the $\
\rho -$dependent (and so flexible) harmonic oscillator potential $\ V_{\text{%
FHO}}(Q_{z}^{+})=\frac{1}{2}\omega _{0}^{2}(\rho )\ Q_{z}^{+\ 2}$ instead$.$
By approximating the ground state of hydrogen by that of the harmonic
oscillator, as is customary in stopping-power theory, we find (see Appendix
2)%
\begin{equation}
\omega _{0}(\rho )=\left( \frac{1}{\rho }\frac{K_{1}(2\rho )}{K_{2}(2\rho )}%
\right) \simeq \frac{1}{\left( 1+\rho \right) },  \label{530}
\end{equation}%
where $K_{n}$ is the modified Bessel function of the second kind. In this
way we obtain the best of both worlds: near the nucleus, $\rho \sim 0,$ $%
\omega _{0}(0)=1$, so the quasiparticle oscillates as the HO ground state
and its energy is simply 1/2; on the other hand, far from the nucleus, $\rho
\gg 1,\omega _{0}\sim 1/\rho \rightarrow 0$, so the classical particle is
nearly free and oscillates at the mercy of the electric field, reproducing
the effect of the free traveling orbital. \ We finally propose the flexible
harmonic oscillator (FHO) distorted wave function%
\begin{equation}
\chi _{i}^{\text{FHO}}(\overrightarrow{Q}^{+},\overrightarrow{P}^{+}|%
\overrightarrow{r},t)=\psi _{i}(\overrightarrow{r}-\overrightarrow{Q}^{+})\
\exp [i\overrightarrow{P}^{+}\cdot \overrightarrow{r}\ -iL_{\text{FHO}%
}^{+}-iE_{i}t]  \label{531}
\end{equation}%
Note that in this case we have an additional dependence on $\rho $, i.e. $%
\overrightarrow{Q}^{+}=\overrightarrow{Q}^{+}(\rho ,t)$, $\overrightarrow{P}%
^{+}=\overrightarrow{P}^{+}(\rho ,t),$ and $\ L_{\text{HO}}^{+}=L_{\text{HO}%
}^{+}(\rho ,t),$ because of Eq.(\ref{530}). We also want to point out the
contrast between the \textit{rigid} time-evolution of the electronic density
\ $\left\vert \chi _{i}^{\text{RTO}}\right\vert ^{2}$ and the $flexible$
motion of $\left\vert \chi _{i}^{\text{FHO}}\right\vert ^{2}$ with $\rho .$

For excitation or ionization processes, the electron is expected to be
loosely bound, making the use of $\chi _{f}^{\text{RTO}-}$ suitable for
describing the final state, and the corresponding matrix-element amplitude is%
\begin{equation}
A_{f}^{\text{FHO}}(t)=\left\langle \chi _{f}^{\text{RTO}-}\left\vert W_{f}^{%
\text{RTO}-}\right\vert \chi _{i}^{\text{FHO}}\right\rangle .  \label{540}
\end{equation}%
It is important to reiterate here that since $\overrightarrow{q}^{-}(t>\tau
)=0,$ the perturbation $W_{f}^{\text{RTO}-}=V_{C}(\overrightarrow{r})-V_{C}(%
\overrightarrow{r}-\overrightarrow{q}^{-})=0,$ and consequently $A^{\text{FHO%
}}(t>\tau )=0.\ $ As we use $\omega _{0}=\omega _{0}(\rho )$, the motion of
the electron density no longer keeps its \textit{rigid} form, varying its
oscillation amplitude with\ $\rho .$ The probability \ $P_{f}^{\text{FHO}%
}=P_{f}^{\text{FHO}}(t=\tau )$ has the same structure as Eq.(\ref{240}).

\section{RESULTS}

Two theoretical methods are presented here:

\begin{itemize}
\item the RTO given by the amplitude (\ref{145}) where both orbitals, the
final and initial ones, move rigidly centered at $\{q^{-}(t),p^{-}(t)\}$ and 
$\{q^{+}(t),p^{+}(t)\},$ satisfying Eqs.(\ref{142}) and (\ref{90}),\ \
respectively, and

\item the FHO$,$ given by the amplitude ( \ref{540}) where the final orbital
moves rigidly centered at $\{q^{-}(t),p^{-}(t)\}$\ satisfying Eq.(\ref{142})
and the initial orbital moves flexibly with $\rho \ $centered at $%
\{Q^{+}(\rho ,t),P^{+}(\rho ,t)\}$ satisfying Eq.(\ref{280}), with\ $%
\overrightarrow{F}_{\text{FHO}}=-\omega _{0}^{2}(\rho )Q^{+}(\rho ,t),$ and
their analytical expressions are presented in Appendix 1.
\end{itemize}

The structure of the code for calculating the RTO and FHO theoretical
methods is straightforward. It involves a two-dimensional numerical integral
over $\rho $ and $z$. In addition, to calculate the amplitude, a further
integration over time is required to obtain the probability, as depicted in
Eq.(\ref{240}). As previously mentioned, all the calculations are performed
using $N_{cy}=4$ and for a symmetric pulse ($\phi =\pi /2)$. In the
following subsections, we explore the behavior of both permitted and
forbidden transitions by varying the photon energy $\omega $ and the
electric field strength $E_{0}$.

In all the figures, $P_{f}^{\ \text{FHO}}$ is depicted as a thick blue line
and $P_{f}^{\ \text{RTO}}$ as a thin red line. For permitted transitions,
the Born approximation (Eq.(\ref{260})) is shown in gray. For forbidden
transitions, where the Born approximation is zero, we instead use the CV
approximation (Eq.(\ref{263})), which represents the simplest high-energy
distorted-wave approximation available for comparison. We will also compare
with the available TDSE results.

\subsubsection{Permitted transitions 1s$\rightarrow np$}

In Figure 1, we present the permitted $1s\rightarrow 2p$ and $1s\rightarrow
3p$ probabilities for $E_{0}=0.01\ $\ and $0.1$ as a function of the laser
frequency $\omega $, ranging from $0.05$ to $2$. Excitation probabilities $%
P_{f}^{\text{RTO}}$ and $P_{f}^{\ \text{FHO}}$ are displayed in all the
figures alongside the first Born approximation. First, we observe that both
theoretical models tend toward the first Born approximation as the quiver
parameter approaches zero, i.e.\ $E_{0}/\omega \rightarrow 0,$ as shown in
Eq.(\ref{250}). A preliminary diagnostic of the quality of the approximation
is the value of the probability: it must be lower than unity. Our FHO
approximation is below unity for almost all the cases studied here, which is
a first indication of its reliability (we find only one region, at $%
E_{0}=0.125$ around $\omega =\omega _{2}=0.375$, where $P_{2p\ }^{\text{FHO}%
} $ exceeds unity by just 9\%).

The structures of the transition probabilities can be interpreted in terms
of the energy gap,%
\begin{equation}
\omega _{n}=\varepsilon _{f}-\varepsilon _{i}=-\frac{1}{2n^{2}}+\frac{1}{2}.
\label{560}
\end{equation}%
The primary peaks at $\omega _{n}$ ($n=2,3,...$) represent the direct
transitions $1s\rightarrow np$ via a simple one-photon absorption, as
described by the first Born approximation. At lower $\omega $ values,
additional secondary peaks are observed at $\omega _{n}/(2j+1)$, where $j$
is an integer. For $E_{0}=0.01$, these secondary peaks are quite prominent
in the RTO model. Such secondary peaks are also evident in the FHO model but
are orders of magnitude smaller. The secondary peak at $\omega _{n}/3$ can
be explained as a three-step transition (three-photon absorption): $%
1s\rightarrow p\rightarrow d\rightarrow np$, which interferes with the
pathway $1s\rightarrow p\rightarrow s\rightarrow np$. Similarly, the
neighboring lower-frequency peaks at $\omega _{n}/5$ can be attributed to
five-photon absorption, with interference across four possible paths, and so
on. In summary, these secondary peaks are associated with an \textit{odd}
number of photon absorptions.

\subsubsection{Forbidden transitions 1s$\rightarrow ns$}

Forbidden transitions $1s\rightarrow 2s$ and $1s\rightarrow 3s$ are
displayed in Figure 2 for $E_{0}=0.01$\ and $0.1$ as a function of $\omega $%
. In these cases, the Born approximation is evidently null; instead, we
present the CV approximation given by the amplitude (\ref{263}) as a
reference. For small values of the electric field, the primary peak at $%
\omega _{n}$ is negligible, as expected. However, a new pattern emerges:
secondary peaks appear at $\omega _{n}/(2j)$, where $j$ is an integer. These
peaks can be attributed to an \textit{even} number of photon absorptions.
The peak at $\omega _{n}/2$ is explained as a $1s-p\rightarrow ns$
transition. Similarly, secondary peaks are observed at $\omega _{n}/4$,
involving four-step processes with interference between two distinct paths: $%
1s\rightarrow p\rightarrow s\rightarrow p\rightarrow 2s,$ and $1s\rightarrow
p\rightarrow d\rightarrow p\rightarrow 2s$. Again, as $E_{0}$ increases and $%
\omega $ decreases, the secondary peaks interfere strongly with each other.

\subsubsection{Highly-forbidden transitions 1s$\rightarrow $nd and 1s$%
\rightarrow $nf}

Highly forbidden transitions $1s\rightarrow 3d$ and $1s\rightarrow 4f$ are
displayed in Figure 3. Probabilities in these cases are significantly
smaller than before. For the $1s\rightarrow 3d\ $\ transition, there is a
well-defined peak at $\omega \sim \omega _{3}/2$, corresponding to the $%
1s\rightarrow p\rightarrow 3d$ \ transition, with no interference. Another
distinct peak can still be identified at $\omega \sim \omega _{3}/4$,
corresponding to the four-photon transition $1s\rightarrow p\rightarrow
d\rightarrow f\rightarrow d$, which interferes with $1s\rightarrow
p\rightarrow s\rightarrow p\rightarrow d.$ Still, for $E_{0}=0.01\ $\ we can
identify a peak at $\omega \sim \omega _{3}/6\ $\ involving a six-photon
transition, but it disappears at $E_{0}=0.1.$

\ For the $1s\rightarrow 4f$ \ transition, a three-photon transition peak is
observed, corresponding to $1s\rightarrow p\rightarrow d\rightarrow 4f$. And
a five-photon transition \ is clearly observed only at\ $E_{0}=0.01.\ $

\subsubsection{Total excitation probabilities}

Figure 4 displays $P_{n}=\sum_{lm}P_{nlm}\ $\ as a function of the frequency 
$\omega $ for $n$=2, 3, and 4 and compared with the TDSE numerical
calculation, for an intermediate value $E_{0}=0.04$. The agreement of the
FHO with the numerical results is surprisingly good for, say, $\omega >0.15$%
, which lends good credibility to the FHO for dealing with excitation in
this range of values. The RTO model gives very large values, as usual, but
converges rapidly to TDSE for $\omega >0.3.$

Figure 5 shows the total excitation probability, defined as $P^{exc}\simeq
\sum_{n}^{n=4}P_{n}$, for $E_{0}=0.01,\ 0.1$, and $1,$ compared against the
first Born approximation, which only involves np-transitions. The first and
most evident observation is that both models converge to the Born
approximation for larger values of $\omega $ at $E_{0}=0.01$ and $0.1$. $%
P_{exc}^{\text{FHO}}$ remains below unity in almost all the cases, as
previously noted.

For the highest electric field strength, $E_{0}=1$, the secondary peaks
became indiscernible, leaving only some remnant of the primary peak. At such
large values of the electric field, the Born approximation at the primary
peak exceeds unity by two orders of magnitude. In this case, neither model
converges to the Born approximation within the displayed range. At this high
value of $E_{0},$ the contribution of the forbidden excitations is very
significant.

\subsubsection{Dependence on the strength of the electric field E$_{0}$}

Figure 6 illustrates the excitation probabilities for the 2p (right panel)
and 2s (left panel) transitions, as functions of the strength of the
electric field $E_{0}$, at the primary peaks: $\omega =\omega _{2}=0.375$
for 2p, and $\omega =\omega _{2}/3=0.125$ for 2s. The permitted 2p
transition clearly displays a structure typical of perturbation theory. When 
$E_{0}\rightarrow 0$, both theories (RTO and FHO) converge to the Born
approximation, which predicts the well-known $E_{0}^{2}$-dependence. As $%
E_{0}$ increases, say $E_{0}\gg 0.1$, both theories deviate from the Born
approximation. In the RTO model, the probability is exaggerated, exceeding
unity. Conversely, the FHO model remains below unity over most of the range
(with the exception observed around $E_{0}=0.125)$. The forbidden 2s
transition at $\omega _{2}/2$, which describes a two-photon absorption
process, behaves as $E_{0}^{4}$ for small values of $E_{0}$, which can be
interpreted precisely as a \textit{product} of perturbative terms, $%
E_{0}^{2}\times E_{0}^{2}$. However, its range of validity is extremely
narrow, say $E_{0}\ll 0.05.$ At higher values of $E_{0}$, the differences
between the RTO and FHO models seem to diminish, with a discrepancy of less
than an order of magnitude. This behavior may be explained by the fact that
the electric field $E(t)$ dominates over the confining potential $V_{FHO}$.
One might guess that the RTO model becomes meaningful as $E_{0}\rightarrow
\infty .$ Following this line of reasoning, we find that the 4f excitation
starts behaving as $E_{0}^{6}$ at $\omega _{4}/3.$

\subsubsection{Shannon entropy}

One can find that the rule of Oppenheimer holds for large values of $n$, $%
i.e.$ $P_{n}^{exc}\propto n^{-3}.$\ However, it is important to study the $l$%
-distribution for a given value of $n$. The Born approximation only allows
the permitted state, $l=1$ , around $\omega =\omega _{n}$, \ but, as we have
observed, there are huge deviations depending on \ $\omega \ $and $E_{0}.$
To understand the $l$-distribution, we find it convenient to resort to the
Shannon entropy\ \cite{Shannon49}. Let us define a normalized and
dimensionless probability, $p_{nl}=P_{nl}^{exc}/P_{n}^{exc}\ $\ with\ $%
P_{n}^{exc}=$ $\sum_{l}p_{nl}^{exc},$ and define the Shannon entropy\ $S_{n}$
as usual%
\begin{equation}
S_{n}=-\sum_{l}p_{nl}\log \ p_{nl}.  \label{600}
\end{equation}%
S$_{n}$ contains all the information we need. The case of \textit{total
certainty}\ (or no surprise) is $p_{nl}^{\delta }=\delta _{l,l0},$ so that $%
S_{n}^{\delta }=0,$ which is the case, for example, of the Born
approximation\ where $p_{nl}=\delta _{l,1}.$ The other extreme corresponds
to the \textit{total uncertainty}, which considers a uniform distribution, $%
i.e.\ $ $p_{nl}^{U}=1/n,$ also called the Harvey distribution, and it
produces $S_{n}^{U}=\log \ n.\ $This expression prompts us to define an
effective number of populated$\ l$-states $N_{n},$ as%
\begin{equation}
N_{n}=\exp (S_{n}).  \label{601}
\end{equation}%
With this definition, we have all we need to visualize the distribution. The
maximum value of $N_{n}$\ corresponds simply to the uniform distribution,$\ $
$N_{n}^{U}=n,$ \ which is the number of $l$-states of the $n$-shell$.\ $\
The minimum value of $N_{n}$\ corresponds to total certainty, \ $%
N_{n}^{\delta }=\exp (S_{n}^{\delta })=1.$ In Figure 7 we show the values of 
$N_{2},\ N_{3},$ and $\ N_{4},$ for $E_{0}=0.01,\ 0.1,$ and $1.0,$ as a
function of $\omega ,$ calculated with the FHO theoretical method$.\ $At\ $%
E_{0}=0.01$, \ permitted transitions are clearly observed at $\omega _{n}\ $%
\ where $N_{n}=1.$ Furthermore, it is so well defined that we can determine
the width of the primary peak\ with accuracy. Secondary peaks are also
clearly observed\ at $\omega _{n}/2$ and so on. But as $\omega \rightarrow
0, $ the different possible transitions \ and their interferences make the
picture close to uniformity. The same occurs for large values of $E_{0}$, to
the point that for $E_{0}=1\ $the singularity of the permitted transition
disappears and the picture approaches uniformity, or equipartition.

\subsection{Ionization}

The ionization probability follows exactly the same procedure, the only
difference being that we need to use the final continuum \ wave function $%
\psi _{\overrightarrow{k}}(\overrightarrow{r})$ in place of the excited
state $\psi _{f}(\overrightarrow{r})$. It is convenient to expand the
continuum wave function in spherical harmonics; \ after integrating over the
angular distribution of the emitted electron, the differential cross section
in terms of the electron energy $E=k^{2}/2$ is simply given by summing over
all angular momenta,%
\begin{equation}
\frac{dP^{ion}}{dE}=\sqrt{2E}\ \sum_{l=0}^{\infty }\ \left\vert \ \
\int_{0}^{\tau}\ dt\ A_{El}^{\ }(t)\ \right\vert ^{2}.
\end{equation}

Figure 8 displays $dP^{ion}/dE$ as a function of the emitted electron energy
for $E_{0}=0.04$ and four different values of $\omega $, compared with the
exact time-dependent Schr\"{o}dinger equation (TDSE) values. While the RTO
definitively overestimates the TDSE result, the FHO model describes the
multiphoton absorption structure quite well, deteriorating as the number of
absorbed photons increases.

Figure 9 shows another comparison for different values of $E_{0}$ and $%
\omega $. Again, one can observe that the FHO model deteriorates as the
number of \ absorbed photons increases, depending here also on the strength
of the electric field $E_{0}$.

\section{CONCLUSIONS AND FUTURE DEVELOPMENTS}

In this article, we present two theoretical models --- \ an impulse
approximation based on the use of rigid traveling orbitals (RTO) and the
flexible harmonic oscillator (FHO) - to describe the excitation of a
hydrogen atom by a symmetric electric pulse. \ Both models account for the
full response of the electron to the electric pulse in the final channel
described by $\chi _{f}^{RTO-}$. \ The RTO model also considers the full
response to the electric field in the initial channel $\chi _{i}^{RTO+}$, \
while the FHO model incorporates the constraint of the Coulomb potential,
approximated by a flexible harmonic oscillator $\chi _{i}^{FHO+}$.

The RTO model is expected to be valid at very large electric fields, while
the FHO model constrains the response to the electric field by introducing a
harmonic oscillator potential, leading to forced-oscillator motions.

The performance of these methods has been tested and compared under various
conditions, such as field frequency $\omega \ $and electric field intensity $%
E_{0}$, for both permitted and forbidden transitions. We find the FHO model
to be robust and reliable, as its probability remains below unity regardless
of the field intensity, and it converges to the Born approximation for
permitted transitions. These methods were extended to ionization processes
by including the continuum state in the final channel. Comparison with
numerical TDSE calculations establishes the range of validity in terms of
the number of photons absorbed. The generalization to multiple pulses is
straightforward. \ Thus, the motion of the classical particle is still
governed by Newton's equation, but now incorporating the sum of the electric
fields. A difficulty arises when the fields oscillate in different
directions, leading to the loss of axial symmetry.

One of the most promising applications of this model appears to be in the
study of molecules. In such cases, the model would require as many classical
particles as there are atoms composing the molecule. These particles
interact not only with the electric field but also with each other,
resulting in coherences, normal modes, and potentially collective
oscillations, which is the goal of this project.

\section{APPENDIX 1. SOLUTION OF THE NEWTON EQUATION}

Consider the following differential equation.%
\begin{equation}
\frac{d^{2}}{dt^{2}}\mathcal{Q}(t)=E(t)-\omega _{0}^{2}\mathcal{Q}(t)=\frac{d%
}{dt}\mathcal{P}(t).  \label{A10}
\end{equation}%
For the FHO\ case, $\omega _{0}$ depends parametrically on $\ \rho ,$\ as
given by Eq.(\ref{530}). Using the Fourier transformation, this equation has
a general solution 
\begin{eqnarray}
\mathcal{Q}(t) &=&\sqrt{\frac{\pi }{2\omega _{0}^{2}}}\operatorname{Re}\left[ i\exp
[+i\omega _{0}t]\ \widetilde{E}(-\infty ,t\ |\ \omega _{0})+i\exp [-i\omega
_{0}t]\ \widetilde{E}(t,\infty \ |\ -\omega _{0})\right] ,  \label{A20} \\
\mathcal{P}(t) &=&-\sqrt{\frac{\pi }{2}}\operatorname{Re}\left[ \exp [+i\omega
_{0}t]\ \widetilde{E}(-\infty ,t\ |\ \omega _{0})-\exp [-i\omega _{0}t]\ 
\widetilde{E}(t,\infty \ |\ -\omega _{0})\right] \ ,  \label{A22}
\end{eqnarray}%
where $\widetilde{E}(t_{1},t_{2}|\omega _{0})$ was defined in Eq.(\ref{262}%
). In our case, the electric field, as given by Eq. (\ref{1}), can be
expanded in a series of six complex exponentials and therefore $\widetilde{E}
$ has a very simple closed form. After a simple unitary rotation, we obtain $%
Q^{+}(t)$ and $P^{+}(t),$%
\begin{eqnarray}
Q^{+}(t) &=&\mathcal{Q}(t)-\mathcal{Q}(0)\cos (\omega _{0}t)-\frac{\mathcal{P%
}(0)}{\omega _{0}}\sin (\omega _{0}t),  \label{A40} \\
P^{+}(t) &=&\mathcal{P}(t)+\mathcal{Q}(0)\sin (\omega _{0}t)\omega _{0}-%
\mathcal{P}(0)\cos (\omega _{0}t),  \label{A50}
\end{eqnarray}%
satisfying $Q^{+}(0)=P^{+}(0)=0.\ $ For $t>\tau ,$\ $E(t>\tau )=0,$ we
should extrapolate the variables as%
\begin{eqnarray}
Q^{+}(t &>&\tau )=Q^{+}(\tau )\cos (\omega _{0}(t-\tau ))+\frac{P^{+}(\tau )%
}{\omega _{0}}\sin (\omega _{0}(t-\tau )),  \label{A89} \\
P^{+}(t &>&\tau )=-Q^{+}(\tau )\sin (\omega _{0}(t-\tau ))\omega
_{0}+P^{+}(\tau )\cos (\omega _{0}(t-\tau )).  \label{A90}
\end{eqnarray}%
This defines the coherent state, which survives, in principle, forever. Its
energy for $t>\tau \ $\ is constant and it is given by 
\begin{equation}
\left. \frac{1}{2}P^{+}(t)^{2}+\frac{1}{2}\omega
_{0}^{2}Q^{+}(t)^{2}\right\vert _{t>\tau }=\frac{1}{2}P^{+}(\tau )^{2}+\frac{%
1}{2}\omega _{0}^{2}Q^{+}(\tau )^{2}.  \label{A100}
\end{equation}

\section{APPENDIX 2. \ DETERMINATION OF THE $\protect\varrho -$DEPENDENT
HOOKE COEFFICIENT.}

Consider the density of the hydrogenic state \ $\psi _{i}(\overrightarrow{r}%
)\ $in cylindrical coordinates $\overrightarrow{r}=(\rho ,z,\phi )$; the
normalized probability as a function of $z$ for a given value of $\rho \ $is%
\begin{equation}
\delta _{i}^{H}(\rho |z)=\frac{\left\vert \psi _{i}(\rho ,z)\right\vert ^{2}%
}{\int_{-\infty }^{\infty }dz\ \left\vert \psi _{i}(\rho ,z)\right\vert ^{2}}
\label{B10}
\end{equation}%
The quadratic moment is simply%
\begin{equation}
\left\langle z^{2}\right\rangle _{i}^{H}(\rho )=\int_{-\infty }^{\infty }dz\
\delta _{i}^{H}(\rho |z)\ z^{2}.  \label{b20}
\end{equation}%
For the ground state of a hydrogenic atom ($Z_{H}=1$ for Hydrogen) we obtain 
\begin{equation}
\left\langle z^{2}\right\rangle _{1s}^{H}(\rho )=\frac{\rho K_{2}(2Z_{H}\rho
)}{2Z_{H}K_{1}(2Z_{H}\rho )}.  \label{b30}
\end{equation}%
Let $\Phi _{n}^{\text{HO}}(\omega _{0}|z)\ $be the quantum solution of the \
Schr\"{o}dinger equation in a harmonic oscillator potential, $%
{\frac12}%
\omega _{0}^{2}z^{2},$ the quadratic moment is%
\begin{equation}
\left\langle z^{2}\right\rangle _{0}^{\text{HO}}(\rho )=\int_{-\infty
}^{\infty }dz\ \left\vert \Phi _{0}^{\text{HO}}(\omega _{0}|z)\right\vert
^{2}\ \ z^{2}=\frac{1}{2\omega _{0}}.  \label{B40}
\end{equation}%
By equalizing $\left\langle z^{2}\right\rangle _{0}^{\text{HO}}(\rho
)=\left\langle z^{2}\right\rangle _{i}^{H}(\rho )$, we determine the
equivalent value of the Hooke coefficient $\omega _{0}=\omega _{0}(\rho )$
as given by 
\begin{equation}
\omega _{0}(\rho )=\frac{Z_{H}K_{1}(2Z_{H}\rho )}{\rho K_{2}(2Z_{H}\rho )}
\label{B50}
\end{equation}

\begin{figure*}[!htb]
\centering
\includegraphics[width=0.90\textwidth]{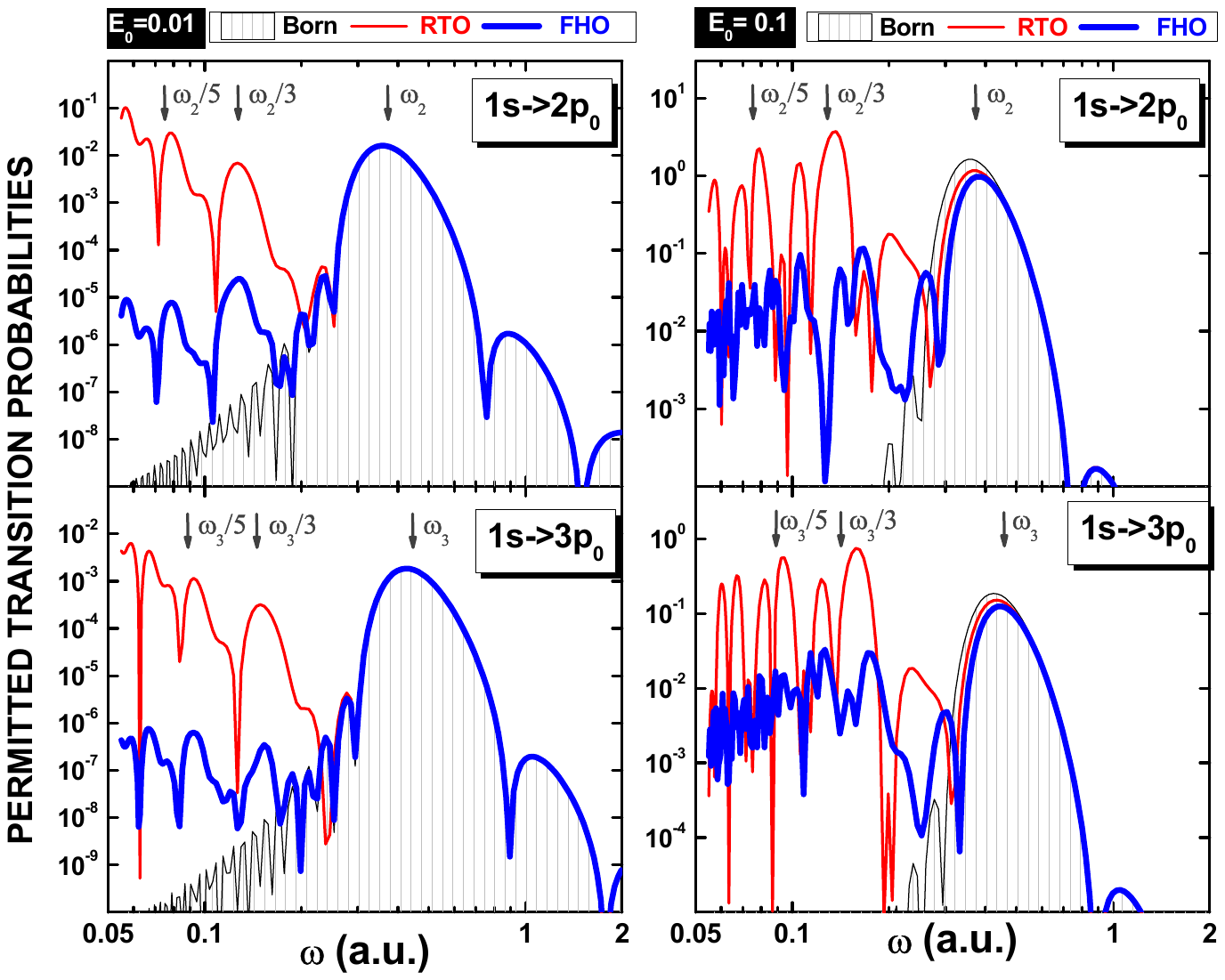}
\caption{(Color online) Transition probabilities for the excitation of
hydrogen from the 1s to 2p state (upper panels) and to 3p state (lower
panels) are shown as a function of the frequency $(\protect\omega)$, for a
symmetric field with $E_0 = 0.01$ (left panels), and $0.1$ (right panels),
as indicated.}
\label{Figure1}
\end{figure*}
\begin{figure*}[!htb]
\centering
\includegraphics[width=0.90\textwidth]{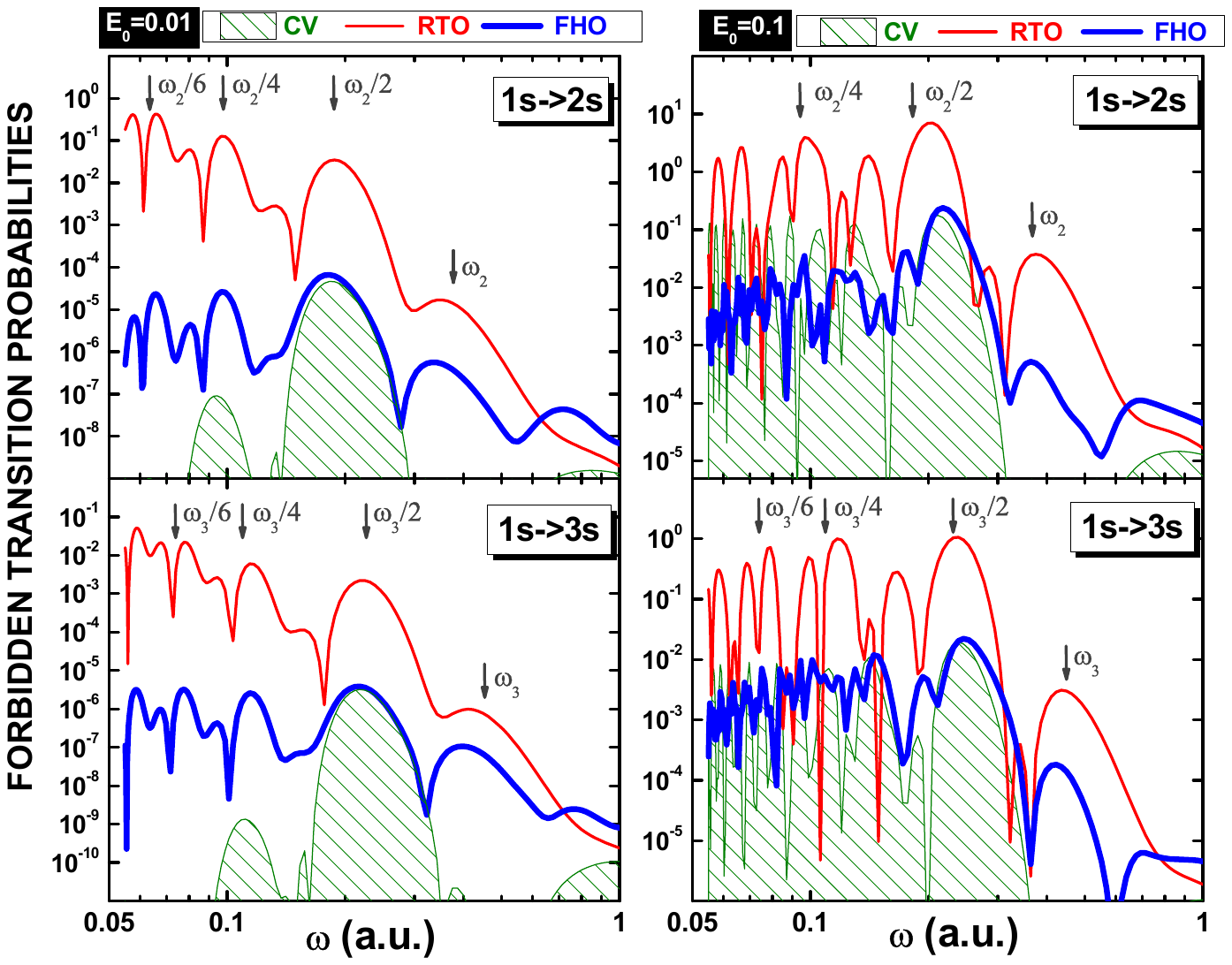}
\caption{(Color online) Transition probabilities for the excitation of
hydrogen from the 1s to 2s state (upper panels) and to 3s state (lower
panels) are shown as a function of the frequency $(\protect\omega)$, for a
symmetric field with $E_0 = 0.01$ (left panels), and $0.1$ (right panels),
as indicated.}
\label{Figure2}
\end{figure*}
\begin{figure*}[!htb]
\centering
\includegraphics[width=0.90\textwidth]{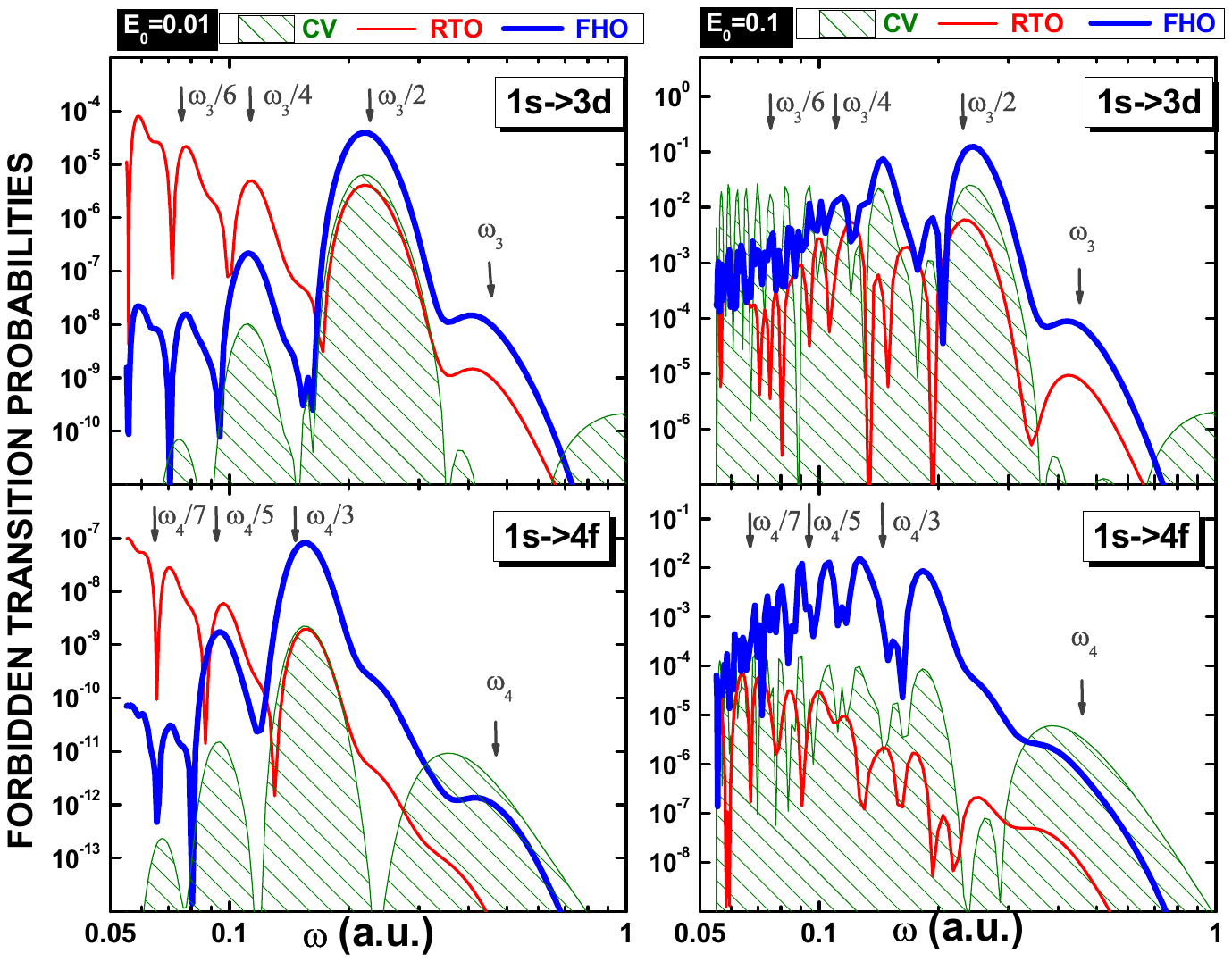}
\caption{(Color online) Transition probabilities for the excitation of
hydrogen from the 1s to 3d state (upper panels) and to 4f state (lower
panels) are shown as a function of the frequency $(\protect\omega)$, for a
symmetric field with $E_0 = 0.01$ (left panels), and $0.1$ (right panels),
as indicated.}
\label{Figure3}
\end{figure*}
%
\begin{figure*}[!htb]
\centering
\includegraphics[width=0.90\textwidth]{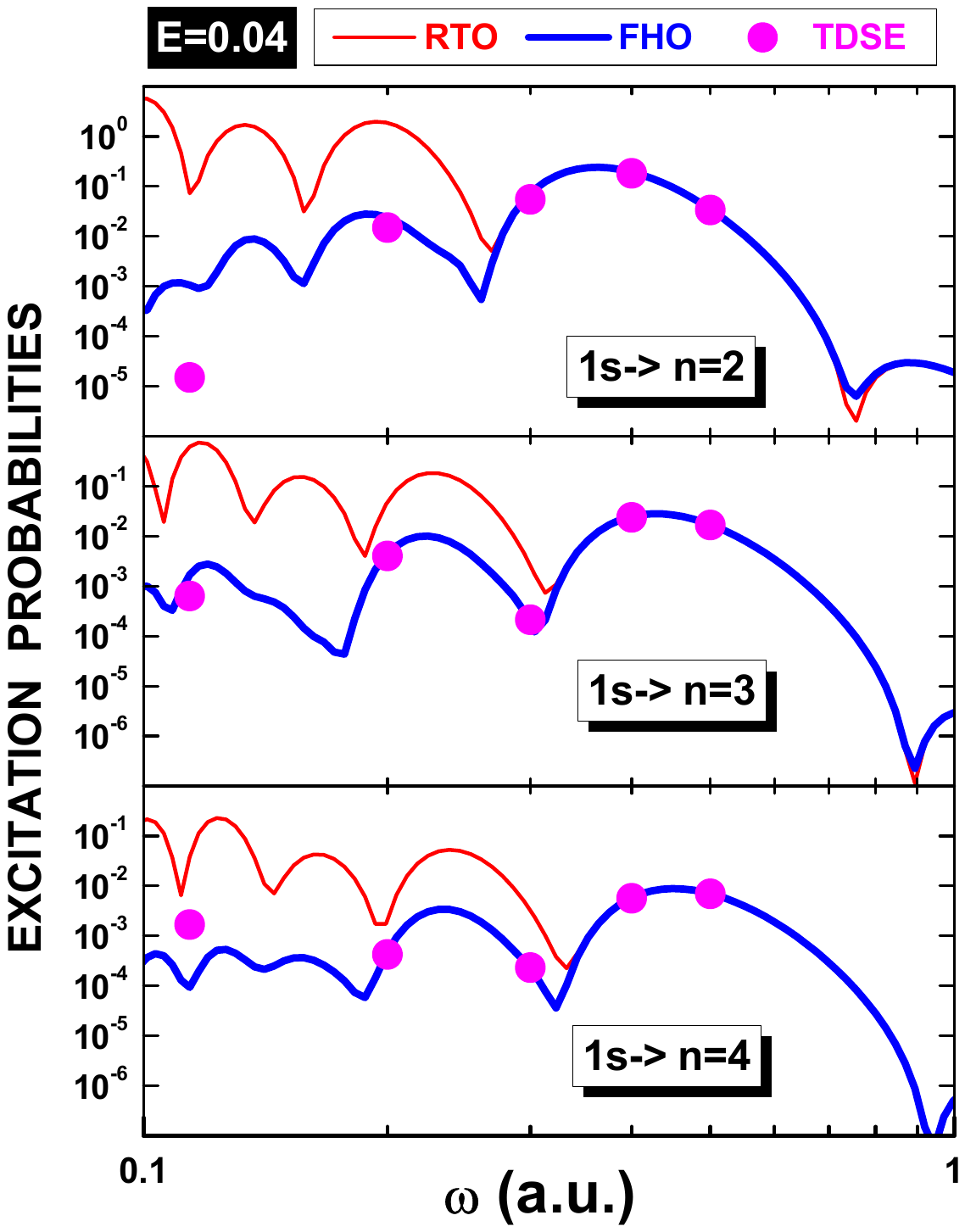}
\caption{(Color online) Transition probabilities for the excitation of
hydrogen from the 1s to n=2, 3 and 4, as indicated, are shown as a function
of the frequency $(\protect\omega)$, for a symmetric field with $E_0 = 0.04$%
. The large circle symbols are the TDSE numerical results.}
\label{Figure4}
\end{figure*}
\begin{figure*}[!htb]
\centering
\includegraphics[width=0.90\textwidth]{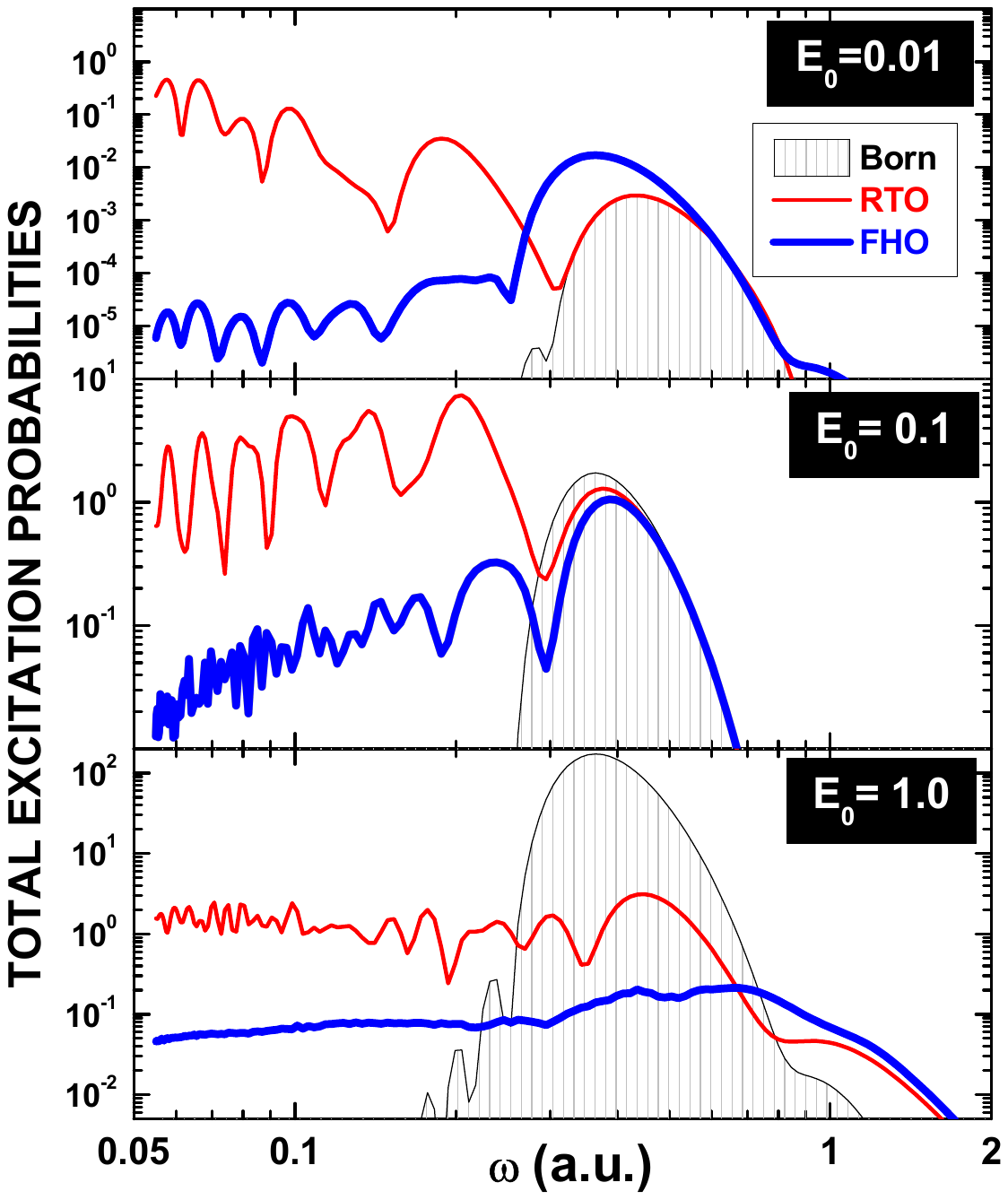}
\caption{(Color online) Total excitation probabilities are displayed for
both models, HO$\protect\rho$ and IA, for $E_{0} = 0.01$ (upper panel), $0.1$
(middle panel), and $1.0$ (lower panel), as a function of the frequency $(%
\protect\omega)$.}
\label{Figure5}
\end{figure*}
\begin{figure*}[!htb]
\centering
\includegraphics[width=0.90\textwidth]{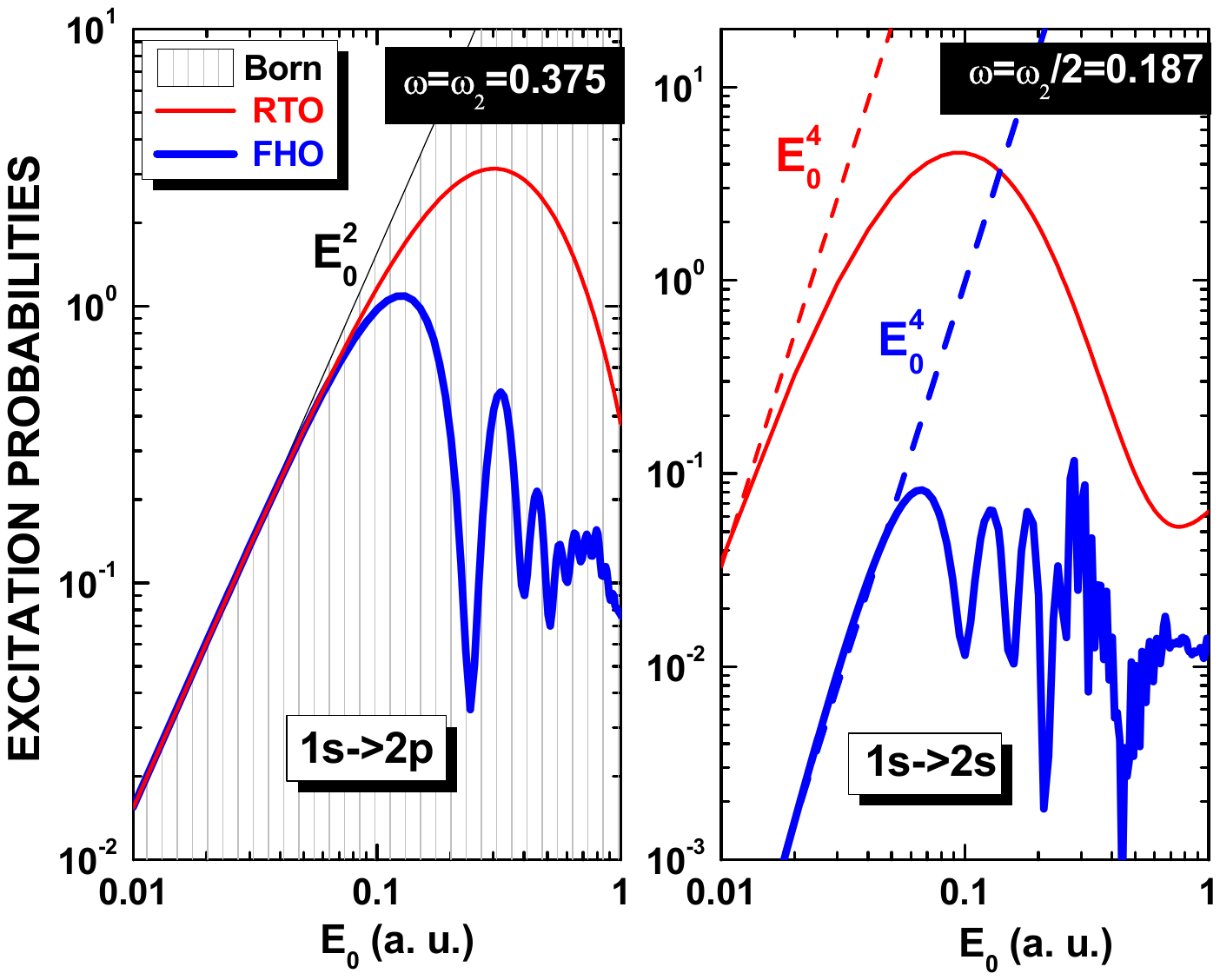}
\caption{(Color online) Left panel: Excitation probabilities for 2p
excitation as a function of the strength of the electric field for $\protect%
\omega_{2}=0.375$. Right panel: Excitation probabilities for 2s excitation
as a function of the strength of the electric field for $\protect\omega%
_{2}/3=0.125.$}
\label{Figure6}
\end{figure*}

\begin{figure*}[!htb]
\centering
\includegraphics[width=0.90\textwidth]{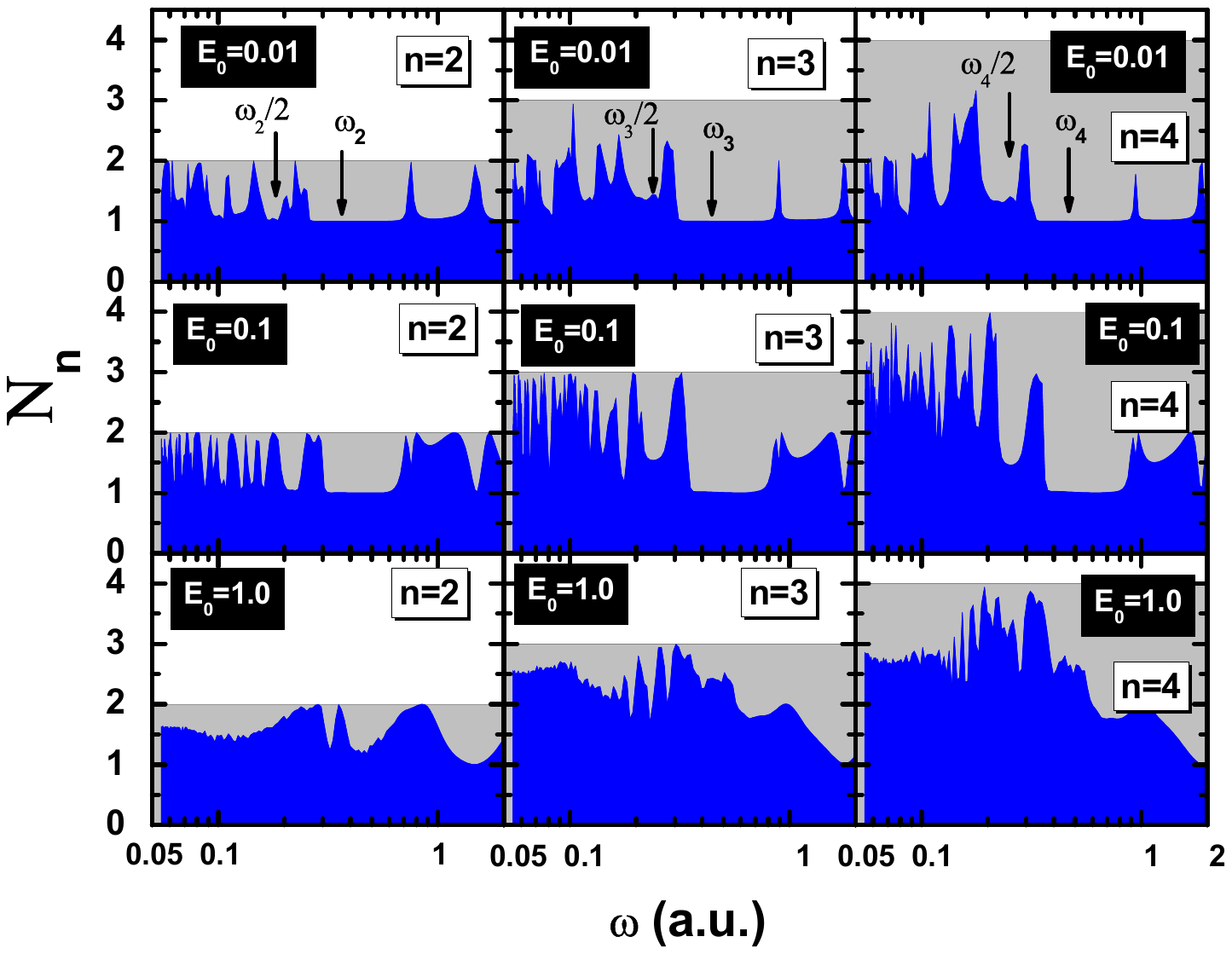}
\caption{(Color online) Effective number of states $N_n$ determined via the
Shannon entropy as a function of the frequency $(\protect\omega)$ for $n=$2,
3 and 4 and $E_0=$0.01, 0.1 and 1.0, as indicated, calculated with the HO$%
\protect\rho$ model. }
\label{Figure7}
\end{figure*}

\begin{figure*}[!htb]
\centering
\includegraphics[width=0.90\textwidth]{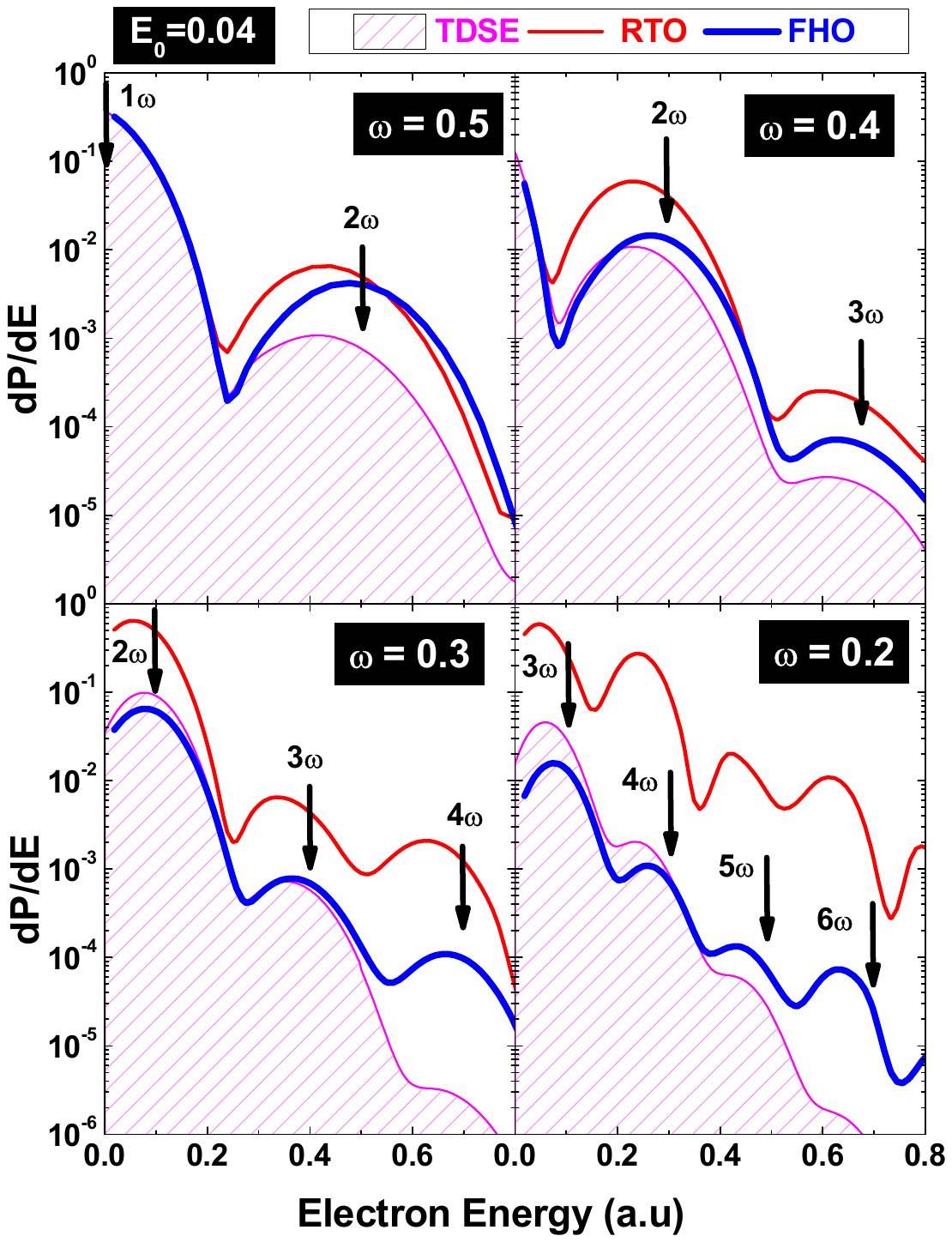}
\caption{(Color online) Energy distributions as a function of the emitted
electron energy for $E_{0} = 0.04$ for four different values of $\protect%
\omega$ and compared with the Time-Dependent Schr\"{o}dinger equation (TDSE)
numerical results, shown with filled circle symbols. }
\label{Figure8}
\end{figure*}
\begin{figure*}[!htb]
\centering
\includegraphics[width=0.90\textwidth]{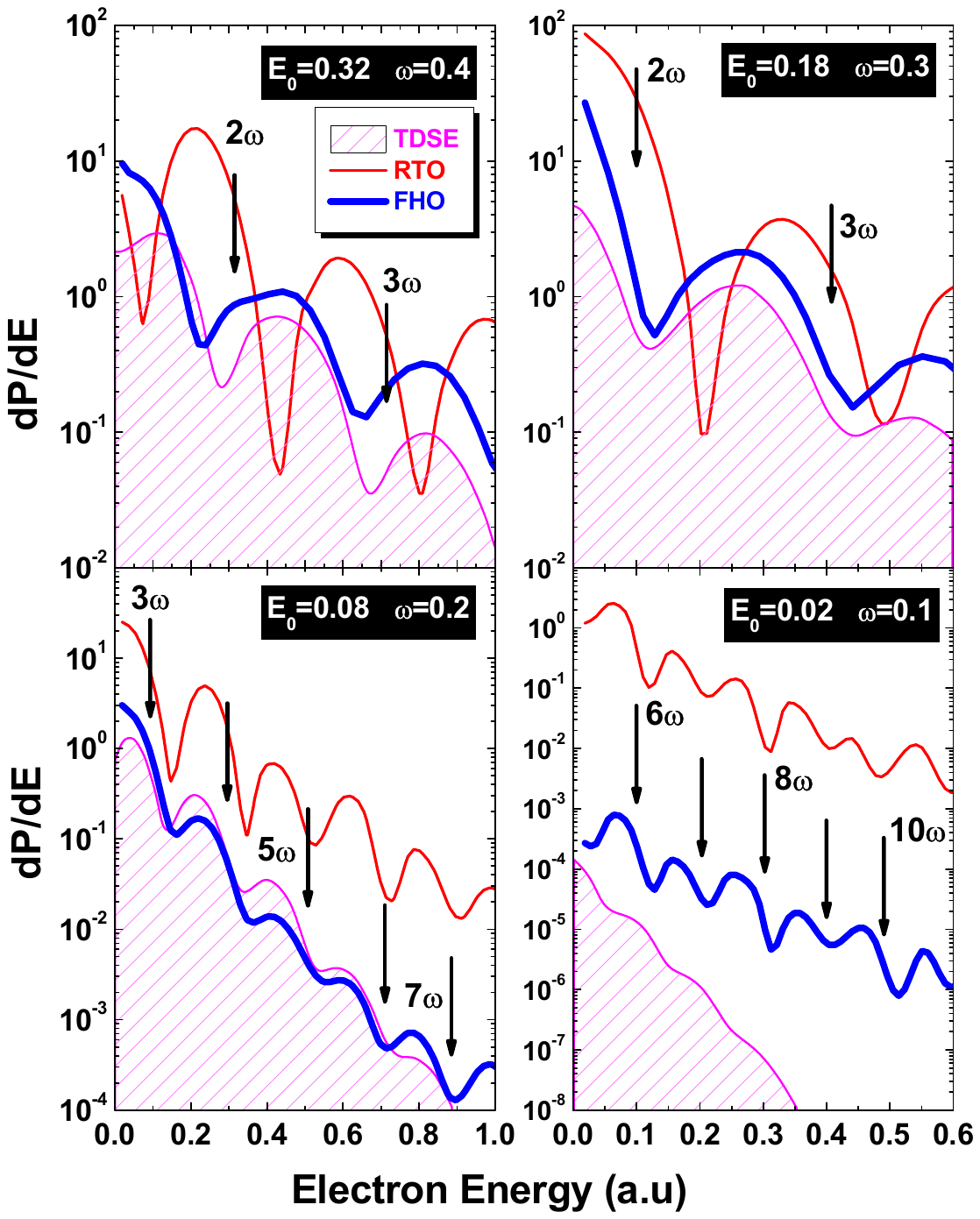}
\caption{(Color online) Energy distributions as a function of the emitted
electron energy for four different values of $\protect\omega$ and $E0$ and
compared with the Time-Dependent Schr\"{o}dinger equation (TDSE) shown with
filled circle symbols. }
\label{Figure9}
\end{figure*}


\end{document}